\documentclass[twocolumn,showpacs,preprintnumbers,amsmath,amssymb]{revtex4}

\usepackage{graphicx}
\usepackage{dcolumn}
\usepackage{bm}

\newcommand{\ket}[1]{\lvert#1\rangle}

\expandafter\ifx\csname
 natexlab\endcsname\relax\def\natexlab#1{#1}\fi
 \expandafter\ifx\csname bibnamefont\endcsname\relax
   \def\bibnamefont#1{#1}\fi
 \expandafter\ifx\csname bibfnamefont\endcsname\relax
   \def\bibfnamefont#1{#1}\fi
 \expandafter\ifx\csname citenamefont\endcsname\relax
   \def\citenamefont#1{#1}\fi
 \expandafter\ifx\csname url\endcsname\relax
   \def\url#1{\texttt{#1}}\fi
 \expandafter\ifx\csname
 urlprefix\endcsname\relax\def\urlprefix{URL }\fi
 \providecommand{\bibinfo}[2]{#2}
 \providecommand{\eprint}[2][]{\url{#2}}

\begin{document}

\title{Backward retrieval in optical quantum memory controlled by an external field}

\author{Alexey Kalachev}
\email{kalachev@kfti.knc.ru} \affiliation{Zavoisky
Physical-Technical Institute of the Russian Academy of Sciences,
Sibirsky Trakt 10/7, Kazan, 420029, Russia
}%

\author{Stefan Kr\"{o}ll}
\email{Stefan.Kroll@fysik.lth.se}
\affiliation{%
Department of Physics, Lund University, Box 118,
S-221 00 Lund, Sweden
}%

\date{\today}

\begin{abstract}
A scheme for backward retrieval in optical quantum memories in which information is stored in collective states of an extended resonant atomic ensemble is developed such that phase conjugation can be implemented by application of an external nonuniform electric (magnetic) field without use of coherent exciting pulses. The possibilities of realizing such a scheme using resonant solid-state materials are discussed.
\end{abstract}

\pacs{42.50.Gy, 32.80.Qk}

\maketitle

In recent years much effort has been directed toward the
implementation of quantum memories for photons. Such devices would
form a basic ingredient for scalable all-optical quantum computers
\cite{KMNRDM_2007} and they are also necessary for quantum
repeaters \cite{BDCZ_1998}, which allow long-distance quantum
communication. Several different approaches to the realization of
optical quantum memories have been proposed on the basis of the
interaction of a single photon with a single atom \cite{CZKM_1997}
as well as with an ensemble of atoms
\cite{FL_2000,KMP_2000,MK_2001,JSCFP_2004,KK_2006}. Among the
latter, quantum memories based on controlled reversible
inhomogeneous broadening (CRIB)
\cite{MK_2001,MTH_2003,ALS_2004,NK_2005,KTG_2005} is of particular
interest from the experimental point of view since it can allow storage
and retrieval without application of additional exciting laser
pulses. Although the CRIB procedure can be realized in gases
using Doppler broadening \cite{MK_2001} and in solids using a
dipole-dipole interaction \cite{MTH_2003}, the simplest
implementation is possibly on the basis of the Stark effect
\cite{ALS_2004,NK_2005,KTG_2005}. Without use of additional laser
fields, the storage time is limited by the phase relaxation time on the
optical transition, which may be of the order of several milliseconds in
rare-earth-ion-doped materials \cite{M_2002}. Proof-of-principle
experimental demonstrations of the CRIB approach with bright
coherent pulses have recently been performed
\cite{ALSM_2006,ALSM_2007,HLALS_2008}.

Regardless of the approach, high efficiency and fidelity of storage and retrieval of information appear to be possible only for optimal conditions
of the light-matter interaction, e.g., an optimal temporal shape of the single-photon wave packet to be stored
\cite{CZKM_1997,GAFSL_2006,GALS_2007,K_2007}. In the case of the interaction of a single photon with extended atomic ensembles, backward information
retrieval has been assumed to be one of the necessary conditions \cite{MK_2001,KTG_2005,GALS_2007}, although recent results indicate that this actually may not
be necessary \cite{HLALS_2008}. In the CRIB approach it is usually assumed that the realization of backward retrieval involves the action of two
additional coherent $\pi$ pulses, which can provide the phase shifts required for creating the coherent emission \cite{MK_2001,NK_2005}. This
circumstance obviously discounts the advantage of the CRIB approach mentioned in the previous paragraph, that the application of additional exciting
laser pulses may be avoided. In the present work we investigate the possibility of backward retrieval by application of an external electric or
magnetic field without any coherent excitation required for creating the coherent emission. It should be noted that the idea to exploit a
longitudinal field gradient for backward retrieval has been proposed in \cite{KTG_2005}. In comparison with that work, we develop a scheme with a
periodic field gradient \cite{Vienna_2005} which makes it possible to realize phase conjugation in an extended optically dense sample for microsecond time delays.
First we show such a possibility following the approach based on storage in subradiant states \cite{KK_2006} and next discuss it in the context
of the CRIB approach.

Consider an extended system of identical two-level atoms forming an optically dense resonant medium. We assume that the atoms are
impurities embedded in a solid-state material and the frequency of the resonant transition can be controlled by application of an external electric or magnetic field. Both possibilities will be discussed below but now we consider the former case for definiteness (Fig.~1).
\begin{figure}
\includegraphics[width=7cm]{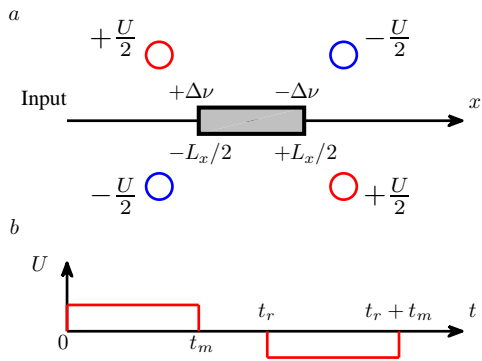}
\caption{(Color online) General scheme of a quantum memory device based on a two-level extended atomic system, collective states of which are controlled by an external electric field (\emph{a}) and time order of application of an external electric field during storage and retrieval via subradiant states (\emph{b}).}
\end{figure}
Suppose that the strength of the field varies linearly along the sample axis $x$ and is oriented perpendicularly to the axis such that the frequency shift varies from some positive value $+\Delta\nu$ at the input end of the sample $x=-L_x/2$ to some negative value $-\Delta\nu$ at the opposite end $x=+L_x/2$. Such a field may be created by four linear electrodes in a quadrupole configuration and was used in proof-of-principle experiments \cite{ALSM_2006,ALSM_2007,HLALS_2008} with CRIB. Now assume that a single-photon wave packet resonant with a
the transition $\ket{0}\to\ket{1}$ propagates through a medium that has a phase relaxation time much longer than the duration of
the wave packet. At some moment of time, say $t=0$, the probability of finding the medium in the excited state and the field in the vacuum state is
at its maximum and at this moment the atomic system can be subjected to an external inhomogeneous electric field, which leads to phase mismatching and inhibition of the collective forward emission. This step corresponds to the writing of information provided that the electric field is sufficiently strong so that phase mismatching may be introduced during a period of time much shorter than the duration of the wave packet being stored. The spatial distribution of the excitation in the medium at $t=0$ depends on the time shape of the input photon and the optical density of the resonant medium \cite{GALS_2007,K_2007}. The spatial distribution is crucial for the quantum memory efficiency, but not for the possibility of backward retrieval discussed here. Let the inhomogeneous electric field be switched on at the moment of time $t=0$. Then, for each time $t_m>0$ satisfying $\Delta\nu t_m=m/2$, where $m$ is a positive integer number, a subradiant state is created \cite{KK_2006}, since for each atom there is an atom opposite in phase with respect to the forward emission, and the rate of collective spontaneous emission in the forward direction proves to be equal to zero. More precisely, at each moment $t_m$, an additional phase shift $\exp(-i\Delta k x)$, where $\Delta k=4\pi\Delta\nu t_m/L_x$, has been created due to the frequency shift. The spatial period of this phase modulation is $\Delta x\sim 1/\Delta k$. It decreases with $m$, and if $m$ is sufficiently large this period will have a thickness corresponding to an optically thin layer $1/\alpha$, where $\alpha$ is the resonant absorption coefficient. In this case the collective spontaneous emission in the forward direction remains totally forbidden in an optically dense medium for all subsequent moments of time and we can switch off the external field. Such subradiant states are able to store the information during a time equal to the phase relaxation time in the medium, $T_2$, and they are equivalent to those created by the switching of phase modulators discussed in Ref.~\cite{KK_2006} when the number of modulators is sufficiently large. In order to read out the information it is necessary to apply the external field with opposite direction for a time $t_m$ again. Then at the moment $t_r+t_m$, where $t_r$ is the time when the field was switched on again (corresponding to the reading step), the subradiant state will be transformed back into the initial superradiant state (i.e., the state created at time $t=0$), and we should switch off the external field in order to allow the atomic system to release the stored excitation in the forward direction as an output photon. Such a regime corresponds to forward retrieval of information, the efficiency of which was analyzed from different points of view in Refs. \cite{GAFSL_2006,GALS_2007,K_2007}. The main result (which basically could be expected) is that the efficiency of backward retrieval may be much better than that of forward retrieval, which also was realized in \cite{MK_2001}, and scales as $1-(\alpha L)^{-1}$, where $L$ is the length of the sample.

Now, let the first external field, which writes the information, remain switched on at the moment $t_m$ and after that. What will happen? The evolution of the spatial phase distribution in the medium may be described as $\exp[ik(t)x]$, where $k(t)=k(0)-\beta t$, $\beta=4\pi\Delta\nu/L_x$, and $k(0)$ is the wave vector of the polarization wave in the medium at time $t=0$. Therefore, when the external field has been applied for a time
\begin{equation}\label{Time}
t_\text{rev}=\frac{1}{\Delta\nu}\frac{L_xn}{\lambda},
\end{equation}
where $\lambda=\nu/c$ is the wavelength of radiation in the vacuum, $\nu$ is the frequency of the resonant transition without an external field, and $n$ is the refractive index of the medium,
we obtain $k(t)=-k(0)$, which corresponds to spatial phase conjugation. If we switch off the external field at this time, the stored information is retrieved in the backward direction.

First of all, it is necessary to estimate whether such a transformation is possible using realistic parameters for the electric field and the storage material. Currently, one of the most commonly discussed solid-state materials for the optical quantum storage are rare-earth-ion-doped crystals \cite{M_2002}, in which the phase relaxation time at cryogenic temperatures may be as long as several milliseconds. In such materials it is possible to prepare narrow absorbing peaks on a nonabsorbing background, i.e., isolated spectral features corresponding to a group of ions absorbing at a specific frequency, using hole-burning techniques and sufficiently narrow laser excitation \cite{PSM_2000,SPMK_2000,NROCK_2002,SLLG_2003,NRKKS_2004,RNKKS_2005}. The peaks can have a width of the order of the homogeneous linewidth, if a laser with a corresponding linewidth is used for the preparation. Rare-earth-ion-doped crystals can be divided into two groups depending on the ions used as impurities. The first group consists of crystals doped by even-electron ions such as $\text{Pr}^{3+}$, $\text{Eu}^{3+}$, $\text{Ho}^{3+}$, and $\text{Tm}^{3+}$ for which the electronic moment is usually quenched and the electronic states are singlets. If such ions are imbedded in  low-spin or nuclear-spin-free hosts, such as $\text{Y}_2\text{SiO}_5$, the phase relaxation time can be hundreds of microseconds also at zero magnetic field and even longer in a nonzero field \cite{M_2002}. In these crystals the electronic transition frequencies can be shifted by the application of an external electric field interacting with the ions via the linear Stark effect \cite{M_2007}. Consider for example the crystal $\text{Y}_2\text{SiO}_5:\text{Pr}^{3+}$. For the ${}^1D_2(1)-{}^3H_4(1)$ transition of ions in site 1 ($\lambda=605.977$~nm), doping concentration of 0.02\% and liquid helium temperatures, we have $T_2=111\,\mu\text{s}$ without magnetic field and $T_2=152\,\mu\text{s}$ in a magnetic field of $77$~G \cite{ECM_1995}. Taking the Stark coefficient as
$111\text{~kHz}/\text{V\,cm}^{-1}$ \cite{GRWM_1997}, the refractive index $n=1.8$, the length of the sample $L_x=1$\,mm, and the external electric field $10^4 \text{ V\,cm}^{-1}$, which is ten times smaller than a typical value of a breakdown electric field, we obtain $t_\text{rev}\approx 2.7\,\mu\text{s}$. Thus this time may be much smaller than the phase relaxation time if the length of the sample is not very long. However, if a longer sample is used in order to have a higher optical density, backward retrieval can still be carried out in the same way by using a sequence of small samples to which such a transformation is applied. Such a periodic structure may also be created in a bulk crystal or a dielectric waveguide using the hole-burning and laser excitation technique \cite{COMB}. The procedure is illustrated in Fig.~2.
\begin{figure}
\includegraphics[width=7cm]{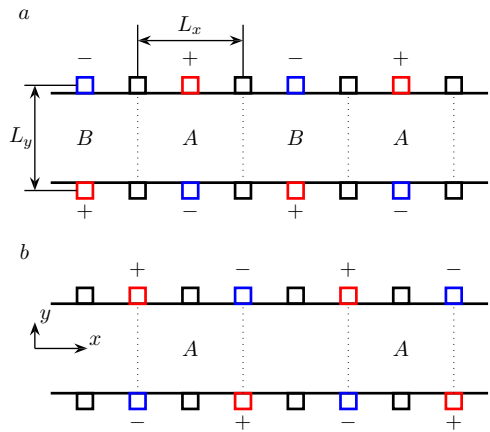}
\caption{(Color online) Scheme for creating the inhomogeneous electric field. The ions need to be removed from the spatial parts $B$ (e.g., by optical pumping) (\emph{a}) and storage will take place in the spatial parts $A$ (\emph{b}). The electrodes are assumed to be infinitely long in the $z$ direction and evaporated on surfaces of a bulk crystal or a dielectric waveguide. The signs near each electrode denote the applied potential $\pm U/2$ while the absence of a sign means $U=0$.}
\end{figure}
First, a narrow absorption peak within a spectral pit is created without an external field. Next a periodical quadrupole field as shown in Fig.~2(a) is applied, such that the ions in spatial regions $A$ and $B$ undergo opposite frequency shifts. Here we assume that the electric dipole moments of the ions are of the same direction, which is perpendicular to the axis $x$. Therefore the dotted lines in Fig.~2(a) show points at which the frequency shift is equal to zero. The applied field is assumed to be not very strong so that the frequency shifts remain within the pit. Under such conditions, ions from one type of the region, say $B$, may be removed by laser excitation. As a result, upon switching off the electric field, we create the necessary periodic structure of impurity ions forming a narrow absorption peak in parts $A$ only. Now, for performing storage and retrieval of information, we should apply a periodic electric field which has the same structure but is shifted along the $x$ axis [Fig.~2(b)]. The resulting system is equivalent to a sequence of small samples placed in the same electric field. It should be noted also that the frequency shifts that are created during the reversal procedure may be much larger than the spectral width of the pit (10--100 MHz). For example, in the case of an $\text{Y}_2\text{SiO}_5$ crystal and $t_\text{rev}\approx 2.7\,\mu\text{s}$ considered above, $\Delta\nu$ is of the order of 1 GHz. But this is not a problem since the frequency shifts are created after absorption of the photon and before its emission, i.e., only during the information storage time.

Now consider the second group of crystals doped by odd-electron ions such as $\text{Nd}^{3+}$ and $\text{Er}^{3+}$. In such materials the phase relaxation time may also be very long but only in the presence of a strong external magnetic field. Moreover, the longest value of $T_2$ in rare-earth-ion-doped crystals reported so far, namely 6.4 ms, is achieved just in $\text{Y}_2\text{SiO}_5:\text{Er}^{3+}$ for the ${}^4I_{15/2}(1)-{}^4I_{13/2}(1)$ transition in a magnetic field of 70~kG \cite{STCEH_2002}. In contrast to the previous group of ions, the odd-electron ions have nonzero electronic magnetic moment in the crystal-field states. This means that we can use a rather small inhomogeneous magnetic field, which adds to a high dc magnetic field to realize the necessary phase transformation. The process of formation of a photon echo in the presence of such magnetic fields was investigated in \cite{WBRM_1990}. Assuming that $d\nu/dH$ is of the order of 1.5~MHz/G, we conclude that the field of 70~G should be applied to realize the transformation in a sequence of 100~$\mu$m samples during 3~$\mu$s. Such a magnetic field can be created in the same way as the electric field considered above, provided that the signs of the potentials are replaced by currents of opposite directions.

The next question that may arise in connection with the suggested approach is whether external fields with sufficient spatial
linearity can be created with real electrode or wire
configurations. We discuss this question for the case of an
electric field, though the same arguments may also be made for the
magnetic field. Let the nonlinearity of the field $E$ as a
function of $x$ be characterized by the root mean square of the
residuals with respect to a linear fit, $\delta E$. We denote the
corresponding quantity in the frequency domain as $\delta\nu$. The
nonlinearity leads to dephasing of atomic states during the
transformation, which is not reversible according to the protocol.
This dephasing will not be destructive if $\delta\nu t_\text{rev} <
1/4$; therefore from Eq.~(\ref{Time}) we obtain the following
constraint:
\begin{equation}\label{nonlinearity1}
\frac{\delta\nu}{\Delta\nu}<\frac{1}{4}\frac{\lambda}{L_xn}.
\end{equation}
Numerics show that the simple geometry depicted in Fig.~2(b) provides
$\delta\nu/\Delta\nu\sim 10^{-2}$ within spatial parts $A$ in the
neighborhood of axis $x$, provided that $L_x\approx L_y$. In other
words, such a configuration is well suited for dielectric
waveguides and implies that $L_x$ is of the order of several
$\lambda$. On the other hand, confining oneself to the linear
dimensions $L_x$,  $L_y\sim 100\lambda$, which is a typical
diameter of an optical fiber cladding, and taking the refractive
index $n=1.8$, we conclude that $\delta\nu/\Delta\nu < 1.4\times
10^{-3}$. To obtain such accuracy we can take advantage of an
approach that is well known in electronic and ionic optics (see,
for example, \cite{S_1988}): the larger the number of electrodes,
the better they can reproduce a given field distribution.
Therefore consider the configuration shown in Fig.~3, in which
twice as many electrodes per spatial region $A$ are involved.
\begin{figure}
\includegraphics[width=7cm]{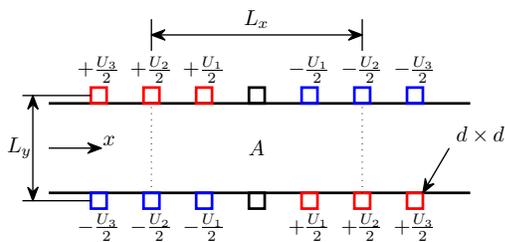}
\caption{(Color online) Configuration of electrodes creating $\delta\nu/\Delta\nu\sim 10^{-4}$ along the $x$ axis within the spatial region~$A$.
}
\end{figure}
Numerics show that, taking the cross-sectional dimension
$d=0.05L_x$, we can achieve $\delta\nu/\Delta\nu < 6\times 10^{-4}$
within a core of $|y|<L_y/10$ when $L_y=0.75L_x$ and
$U_1:U_2:U_3=0.518:1:2.14$. The maximum value of the 
electric field created at the boundaries of $A$ proves to be
$|E_\text{max}|\approx U_2/L_y$. If we consider
$\text{Y}_2\text{SiO}_5:\text{Pr}^{3+}$ again and assume
$L_y=100\lambda=60.6$~$\mu$m and $|U_2/2|=5$~V, we obtain
$\Delta\nu=183$~MHz and $t_\text{rev}=1.3$~$\mu$s. From the
relation $\Delta\nu t_m=m/2$ it can be seen that different
orthogonal subradiant states alternate with each other at the
interval $t_{m+1}-t_m=1/2\Delta\nu$, which in the considered
example proves to be equal to 2.7~ns. The accuracy with which the
electric field has to be switched on and off in time should be
smaller than $\frac{1}{4}(t_{m+1}-t_m)$. On the other hand, the
external field need not be a rectangular function of time: it may
be strong in the middle of the transformation and weak at the
onset and completion of it, provided that $t_{m+1}-t_m$ remains
smaller than the wave packet duration. In this case the total
change of an atomic phase $\int\Delta\nu(t)\,dt$ only should be
controlled so that $\overline{\Delta\nu} t_\text{rev}<1/4$, where
$\overline{\Delta\nu}$ is the time average of the deviation of
$\Delta\nu(t)$ from a given dependence on time.

The dephasing of atomic states due to the nonlinearity of the
field reduces the efficiency of backward retrieval by the factor
of $\cos^2(2\pi\delta\nu t_\text{rev})$. From this point of view
the condition (\ref{nonlinearity1}) should be replaced by
\begin{equation}\label{nonlinearity2}
\frac{\delta\nu}{\Delta\nu}<\frac{\arccos(\sqrt{\epsilon})}{2\pi}\frac{\lambda}{L_xn},
\end{equation}
where $\epsilon$ is required efficiency. The considered example ($\delta\nu/\Delta\nu = 6\times 10^{-4}$, $L_x=133\lambda$, $n=1.8$) corresponds to
$\epsilon=0.38$. The efficiency of $\epsilon=0.9$ (0.99) requires $\delta\nu/\Delta\nu < 2.14\times 10^{-4}$ ($6.7\times 10^{-5}$), which needs more electrodes per spatial region. A further increase of that number from 8 to 12 makes it possible to reach $\delta\nu/\Delta\nu< 5\times 10^{-5}$,
which is sufficient for high efficiency.

Finally, we note that the developed scheme can be readily combined with the CRIB approach, which allows backward retrieval in the time-domain optical quantum memory without the application of additional laser pulses. Between the processes of writing and readout through an inhomogeneously broadened transition the information can be stored on a narrow absorption peak, when a writing or reading external electric field is turned off, and the transformation $k(t)\to -k(0)$ may be implemented during the storage time. In this case the onset (completion) of the transformation does not cause the photon to be absorbed (emitted), so that $t_{m+1}-t_m$ need not be smaller than the wave packet duration. Further, since the backward retrieval does not involve reversal of inhomogeneous broadening, the developed scheme may be considered as a generalization of CRIB approach to controlled inhomogeneous broadening.

A.K. thanks the Department of Physics, Lund University, for hospitality. This work was supported by the
Program of the Presidium of RAS "Quantum macrophysics" and the Royal Swedish Academy of Sciences, the European Commission through the integrated project QAP under the IST directorate, and by the Swedish Research Council.


\end{document}